\documentclass[11pt,a4paper]{article}
\pdfoutput=1
\usepackage{jcappub}
\usepackage{subfig}
\usepackage{subfloat}
\usepackage{float}

\usepackage[english]{babel}
\usepackage{xspace} 
\usepackage{url}

\usepackage{cosmodefs.JCAP}

\usepackage{amsmath}
\usepackage{amssymb}
\usepackage{ifthen,xcolor}
\begin{document}

\date{\today}
\title{On the Unlikeliness of Multi-Field Inflation: Bounded Random Potentials and our Vacuum}
\author{Diana Battefeld,} 
\emailAdd{dbattefe(at)astro.physik.uni-goettingen.de}
\author{Thorsten Battefeld,}
\emailAdd{tbattefe(at)astro.physik.uni-goettingen.de}
\author{Sebastian Schulz}
\emailAdd{sschulz(at)astro.physik.uni-goettingen.de}
\affiliation{Institut of Astrophysics, University of Goettingen, Friedrich-Hund-Platz 1, 37077 Goettingen, Germany}

\abstract{Based on random matrix theory, we compute the likelihood of saddles and minima in a class of random potentials that are softly bounded from above and below, as required for the validity of low energy effective theories. Imposing this bound leads to a random mass matrix with non-zero mean of its entries. If the dimensionality of field-space is large, inflation is rare, taking place near a saddle point (if at all), since saddles are more likely
than minima or maxima for common values of the  potential. Due to the boundedness of the potential, the latter become more ubiquitous for rare low/large values respectively. Based on the observation of a positive cosmological constant, we conclude that the dimensionality of field-space after (and most likely during) inflation has to be low if no anthropic arguments are invoked, since the alternative, encountering a metastable deSitter vacuum by chance, is extremely unlikely.  } 
\keywords{Random Potentials, Multi-Field Inflation, Vacuum Selection}

\maketitle

\section{Introduction}

What is the best model of inflation? Whether or not a model is natural, compelling or beautiful is often in the eye of the beholder, but ultimately consistency with observations is the deciding factor. If a single observation is at odds with a model, it should be abandoned.

Models of inflation in string theory are varied, see \cite{HenryTye:2006uv,Cline:2006hu,Burgess:2007pz,McAllister:2007bg,Baumann:2009ni,Mazumdar:2010sa} for reviews, involving moduli fields, branes and fluxes among other ingredients. One of the first well studied models in string theory is the KKLMMT \cite{Kachru:2003sx}
brane inflation \cite{dvali-tye,Alexander:2001ks,collection,Dvali:2001fw, Firouzjahi:2003zy,Burgess:2004kv,Buchel,Iizuka:2004ct} setup, where  all but one effective degree of freedom are carefully stabilized. However, soon it was realized that in the absence of a compelling a priori reason, several dynamical degrees of freedom appear naturally, leading to a plethora of multi-field inflationary models, see i.e.~\cite{Dimopoulos:2005ac,Piao:2002vf,Majumdar:2003kd,Becker:2005sg,Ashoorioon:2006wc,Ashoorioon:2008qr,Cline:2005ty,Ashoorioon:2009wa,Ashoorioon:2009sr,Battefeld:2010rf,Firouzjahi:2010ga,  Tye:2009ff,Tye:2008ef,Battefeld:2008py,Battefeld:2008qg}
for a small, non-representative selection.

Unfortunately, predictions for correlation functions in the cosmic-microwave-background radiation are often degenerate, offering little discriminating power \cite{Mortonson:2010er,Easther:2011yq,Norena:2012rs,Martin:2010hh,Hazra:2012yn,Easson:2010uw,Easson:2010zy} so far. Non-Gaussianities (NG) can provide such a discriminator if it were observed, since simple single-field models are unable to generate them \cite{Maldacena:2002vr}; however, there is no compelling reason why NG should be large in multi-field models either \cite{Battefeld:2006sz,Battefeld:2007en}: to (temporarily) boost non-linearity parameters, sudden events are required, such as a sharp turn of the trajectory, a sudden (temporary) increase of slow roll parameters, a decrease of the speed of sound, or particle production events, to name a few\footnote{To make accurate predictions for i.e.~fluctuations in the CMBR, one has to follow perturbations until an adiabatic (i.e.~single-field) regime is reached \cite{Battefeld:2009ym,Elliston:2011dr}.}. Although such events can occur, and they might be natural for certain model classes, it is not clear how generic they are.

This uncertainty is compounded by our limited understanding of the landscape in string theory \cite{Susskind:2003kw}. We know little about the origin of the low-energy, effective-field-theory potential in our universe; even if we understood all of the relevant physics, the potential would most likely be far from simple, involving many parameters whose values we would be unable to compute from first principles. An example is the Denev-Douglas landscape \cite{Denef:2004cf}, which was modelled recently  as a random potential \cite{Marsh:2011aa} (see also \cite{Chen:2011ac}).

Fortunately, if a potential of dimensionality $D\gg 1$ is sufficiently complex, some properties, such as the distribution of minima or saddle points, can be predicted quite accurately using random matrix theory (see i.e.~\cite{Mehta:1991,Rao:2005} for a textbook introduction) based on the feature of universality \cite{Deift:2006,Kuijlaars:2011,Bai,Soshnikov:2002,Schenker:2005}.

We follow this approach with the aim to answer a simple question:
\vspace{0.3cm}
\begin{center}
 \emph{What is the inflatons' most likely resting place, if inflation is driven by multiple fields in a random potential that is softly bounded from above and below, with equal likelihood for positive and negative values?} 
\end{center}
\vspace{0.3cm}
The  quantitative answer needs to be compared with the observation of a small, but \emph{positive} cosmological constant (CC) in our universe. 

The novel feature in our analysis is the incorporation of upper and lower bounds, as needed for the viability of effective-field-theory, by means of a Gaussian suppression. As the dimensionality of field-space increases, we find an extraordinarily strong preference of negative values; consequently, almost all inflationary trajectories end up in an anti-deSitter (ADS) minimum, at odds with observations.  In this sense, multi-field inflationary models of this type are strongly disfavored, the more so the higher the dimensionality of field-space $D$ ($D\sim 500$ on the string landscape). While completing this study, a paper by Chen.et.al.~\cite{Chen:2011ac} came to similar conclusions, but the full ramifications for multi-field models were not spelled out. 

One could resort to anthropic reasoning to reconcile this inconvenient observation with multi-field models -- after all, eternal inflation appears to be able to populate all vacua in a landscape \cite{Brown:2011ry} and one of them should be just right for us\footnote{Note that this line of reasoning opens up a whole can of worms with regard to the measure, which is ill defined in an eternally inflating Universe \cite{Olum:2012bn}.}. We refrain from such a reasoning throughout most of this paper, primarily due to our personal bias: anthropic reasoning had been invoked repeatedly in science in the past, oftentimes precluding much simpler explanations. Further, where it is applicable, anthropic reasoning seldom explains anything beyond the observation it was invoked for in the first place -- it primarily puts one's mind at ease. Weinberg's prediction of an upper bound on a positive CC based on anthropics (merely requiring the formation of gravitationally bound objects) consistent with current observations is a notable and admittedly relevant exception \cite{Weinberg:1987dv}. Similar lower bounds for a negative CC exist \cite{Bousso:2008bu}, which are, on the one hand, weaker since a negative CC actually enhances the build-up of structures (on the other hand, a negative CC leads to a re-collapse of the universe).

With this caveat in mind, we take as an agnostic working hypothesis that the CC is not entirely set by anthropics. If it turns out that almost all conceivable models are inconsistent with a positive CC, one may revert back to anthropic reasoning, see Sec.~\ref{sec:anthropics}, but it appears to us that there is no lack of (single or few-field) inflationary setups with the ability to accommodate deSitter vacua\footnote{We do not aim to address the cosmological constant problem here, that is why the magnitude of the CC is so small compared to predictions in the standard model of particle physics; we merely point out that the prediction in a class of multi-field inflationary models is a large negative contribution to the CC.} that avoid the additional challenge brought forth by a large $D$ during inflation.

In addition to drawing conclusions about the final state after inflation (theoretically and numerically), we investigate the feasibility of inflation in random potentials in the first place, following prior research by Tegmark \cite{Tegmark:2004qd} as well as Frazer and Liddle \cite{Frazer:2011tg,Frazer:2011br}. We find that inflation is unlikely (the more so the higher $D$ is), and if it occurs, it almost always takes place near a saddle point. Further, we observe that a short burst of inflation is more likely than a longer one, $P\propto N^{-{\beta}}$, where $N$ is the number of efolds and $\beta$ a model specific parameter of order $\beta \sim 3$, based on the numerical runs in a particularly simple (from a computational point of view) random potential \cite{Frazer:2011tg,Frazer:2011br}, that retains the feature of being softly bound from above and below, recovering and extending results by Frazer and Liddle \cite{Frazer:2011br}; this result is consistent with the theoretical prediction $\beta=3$ for inflation at a saddle point that was reached in \cite{Freivogel:2005vv,Agarwal:2011wm}. In essence, we provide additional theoretical explanations for some of the observations made in \cite{Frazer:2011br}. We would like to point out that many of the theoretical conclusions were reached before, albeit scattered over several publications and years \footnote{Random matrix theory is a well studied branch of mathematics with many applications in i.e.~solid state physics; one of the first applications in cosmology was in the less known, but excellent paper by Azami and Esther \cite{Aazami:2005jf}, which has recently caught the attention of Marsh et.al.~\cite{Marsh:2011aa} and Chen et.al.~\cite{Chen:2011ac}. The preference of shorter bursts of inflation near a saddle point over longer ones goes back to \cite{Freivogel:2005vv} (see also \cite{Agarwal:2011wm}).}.  

The outline of this paper is as follows: to investigate the distribution of minima, maxima and saddle points, we first provide a brief review of known results in random matrix theory, Sec.~\ref{sec:review}. Random matrix theory is needed, because the Hessian is, to a good approximation, a random matrix if enough random parameters enter the low energy effective potential. As a concrete test-bed, we employ a simple random potential retaining the feature of boundedness in Sec.~\ref{sec:softboundpot}. This potential is chosen to be relatively smooth to support inflation. Aided by this toy model, we draw general conclusions about the final state after inflation, Sec.\ref{sec:hessian} and Sec.\ref{sec:minima}, based on universality in random matrix theory. We compare our analytic estimates with numerics in Sec.~\ref{sec:minima}, paying particular attention to the dependence of the minimas probability distribution on the dimensionality of field-space. We investigate the feasibility of inflation in Sec.~\ref{sec:inflation}, again comparing theoretical expectations with numerical results in the toy-model. The consequences for the final state after inflation and the resulting unlikeliness of multi-field inflation are elaborated on in Sec.~\ref{sec:cc}; in Sec.~\ref{sec:anthropics} we comment on a possible way to avoid these conclusions if anthropic arguments in an eternally inflating universe are invoked. We conclude in Sec.~\ref{sec:conclusion}.  
 
\section{Some Properties of Random Potentials and Inflation therein \label{sec:review}}
Here, we present some known results based on random matrix theory in the realm of inflationary cosmology. Readers familiar with this topic may jump to Sec.~\ref{sec:softboundpot}.

\subsection{Likelihood of Saddle-Points VS.~Maxima/Minima \label{sec:review1}}
In \cite{Aazami:2005jf,Marsh:2011aa} random potentials with multiple degrees of freedom and their inflationary trajectories were studied with the aim to extract statistical properties of the distribution of minima, maxima and saddle-points (see also \cite{Chen:2011ac}), as well as the feasibility of inflation on such a landscape. Here, we would like to briefly review some of their main results, before extending them. For a general introduction to random matrix theory see i.e.\cite{Mehta:1991,Rao:2005}.

Consider $D$ scalar fields $\varphi_i$, $i=1,\dots,D$, with potential $V(\varphi_1,\dots,\varphi_D)$, canonical kinetic terms and a flat field-space metric as a model for the landscape in string theory \cite{Susskind:2003kw}. If $D$ is large, a box with volume $(2\Delta)^D$ with $\Delta \lesssim M_p$ is expected  to possess a large number of critical points with $\nabla V=0$ (here $\nabla$ is the gradient in field-space), say $\alpha^D$ with $\alpha \sim \mbox{few}$, and a smaller but still large number of extrema\footnote{By considering a finite box, we are implicitly imposing a measure on the landscape; computing the frequency of saddles to minima/maxima is ill defined on an infinite field-space without imposing further restrictions.}. The proportion of extrema to saddle points is important for several reasons: first, anthropic arguments to explain the smallness of the cosmological constant rely on the presence of many minima ($\sim 10^{500}$) besides a dynamical mechanism to scan this landscape. Second, to drive small field inflation, a close encounter with a saddle-point is needed before the fields get trapped in a minimum with small but positive energy (large field models suffer from the $\eta$ problem, that is quantum corrections usually spoil the needed flatness for super-Planckian field excursions (such corrections are one of the main motivations to consider random potentials); further, inflation driven in a metastable minimum leads to an observationally ruled out open universe after tunnelling in the absence of a subsequent slow roll phase.  Thus, one should expect that inflation near a saddle point is also the most likely occurrence as $D$ is increased in a random potential.)  

Extrema are classified by the Hessian, or mass-matrix, 
\begin{eqnarray} H
= \left(\begin{matrix}
\frac{\partial^2V}{\partial\varphi_1^2} & \ldots & \frac{\partial^2V}{\partial\varphi_1\partial\varphi_D} \\
\vdots & \ddots & \vdots \\
\frac{\partial^2V}{\partial\varphi_D\partial\varphi_1}  &\ldots& \frac{\partial^2V}{\partial\varphi_D^2} 
\end{matrix}\right)\,,
\end{eqnarray}
and if $V$ is a random potential, the Hessian becomes a random matrix. However, in general the entries are not devoid of structure since the potential satisfies constraints; for example, in the super-gravity models of Denef and Douglas \cite{Denef:2004cf}, both the Kaehler- and superpotential are ``random'', leading to a Hessian that can be approximated as a sum of Wigner and Wishart matrices \cite{Marsh:2011aa} (see also \cite{Chen:2011ac}). In \cite{Aazami:2005jf}, a simpler case was studied from a pure phenomenological viewpoint, assuming the Hessian to be a real Wigner matrix (symmetric, Gaussian distributed entries with identical variance and zero mean). Given a Hessian, one needs to find the probability for $H$ to be positive definite: if the Hessian has only positive/negative eigenvalues, a local minimum/maximum is present, and for mixed eigenvalues the critical point is a saddle point.

To compute the probability of a minimum, one needs to derive the joint probability distribution of eigenvalues, followed by an investigation of fluctuations of eigenvalues to compute the probability that the lowest eigenvalue is positive. Performing this task for small fluctuations of the smallest/largest eigenvalues goes back to Tracy and Widom \cite{Tracy:1994} (see also \cite{Forrester:1993}, and \cite{Johansson:2000,Johnstone:2001} for extensions), while large fluctuations, which are needed in our case, have been the subject of a series of papers by Majumdar et.al. \cite{Dean:2006,Vivo:2007,Dean:2008,Vergassola:2008,Nadal:2009,Nada:2011}, see \cite{Marsh:2011aa} for a brief review. The eigenvalue spectrum and fluctuations of extreme eigenvalues turn out to become independent of the statistical properties of the matrix elements in the large $D$ limit \cite{Deift:2006,Kuijlaars:2011,Bai,Soshnikov:2002}, and it appears that this universality still holds if entries are highly correlated \cite{Schenker:2005}. Motivated by this property, we focus on simple toy models in this paper, anticipating that conclusions are true more generally.  

Let us start with the setup of \cite{Aazami:2005jf}, focussing on two extreme cases: first, assume that cross couplings between the fields are small so that the Hessian at the critical point can be diagonalized by a field-redefinition. Second, assume that cross-couplings are large so that the Hessian is a random matrix with elements of comparable magnitude. 

Near a critical point the potential can be written as
\begin{equation}
V(\varphi)=\sum_if_i(\varphi_i)+\sum_{i\neq j}\epsilon_{ij}\varphi_i\varphi_j\,,
\end{equation}
where the $f_i$ each have at least one extremum  ($\partial f_i/\partial \varphi_i=0$), and $\epsilon_{ij}\varphi_i\varphi_j$ parametrize cross-couplings.
In the first (unlikely) case where  the cross couplings, $\epsilon_{ij}$, are small compared to the diagonal elements in $H$, the Hessian is diagonal to a high degree. The eigenvalues are then independent random variables with zero mean and some variance. Due to universality, the matrix elements are taken to be independent Gaussian variables in \cite{Aazami:2005jf}; the probability of a minimum/maximum becomes
\begin{equation}
P= \frac{1}{2^D}\,.\label{psaddle1}
\end{equation}
In other words, the likelihood of a large fluctuation shifting all the eigenvalues to the same side of zero is $1/2^D$. 

If, on the other hand, the cross couplings are of the same order as the diagonal elements, the joint probability distribution obeys Wigners semi-circle law for the Gaussian ensemble \cite{Mehta:1991} and the probability of a minimum/maximum scales as 
\begin{equation}
P\propto e^{-cD^p}\,,
\end{equation} 
with constants $c\approx 1/4$ and $p=2$ \cite{Aazami:2005jf}. Intermediate $\epsilon_{ij}$ interpolate between the two extremes; for instance, decomposing $H=A+B$, where $A$ is diagonal with variance $\sigma_A$ of its Gaussian distributed entries and $B$ random with variance $\sigma_B$, it was found numerically that \cite{Chen:2011ac} 
\begin{eqnarray}
P\propto e^{-c_2D^2-c_1D} \label{mixedHessian}
\end{eqnarray}
where 
\begin{eqnarray}
c_2&\approx& 0.000395y+1.05y^2-2.39 y^3 +\mathcal{O}(y^4)\\
c_2/c_1&\approx& 0.0120+2.99y-12.2y^2+1650y^3+\mathcal{O}(y^4)
\end{eqnarray} and $y\equiv \sigma_A/\sigma_B\ll 0.1$. In general, the constants also depend on the joint probability distribution of the eigenvalues. For complex Wigner matrices one finds $c=ln(3)/2$ and $p=2$ \cite{Aazami:2005jf,Dean:2006,Dean:2008}, while the more complicated Denev-Douglas landscape leads to $c\approx 0.08$ and $p\approx 1.3$ \cite{Marsh:2011aa}. 

The main lesson is that large fluctuations of eigenvalues, as needed for a positive definite Hessian, are extremely rare if the joint probability distribution is reminiscent of Wigner's semi-circle law, much more so than the naive intuition based on statistically independent eigenvalues suggests, $\exp(-cD^p)$ with $1<p \lesssim 2$ vs. $1/2^D$. Consequently, almost all critical points are saddle points.

An intuitive visualization of the joint probability density of the Wigner ensemble in terms of a Coulomb gas is due to Dyson \cite{Dyson:1962.1,Dyson:1962.2}: the eigenvalues correspond to the positions of $D$ charged beads undergoing Brownian motion on an infinite string, with a confining quadratic potential at the origin and logarithmic repulsion between the beads. The probability of finding only positive/negative eigenvalues corresponds to the probability of finding all beads to the left/right of the origin. To shift the outermost bead, a displacement of order $\mathcal{O}(\sqrt{D})$ is needed. Since the potential is quadratic, such a displacement comes at an energy cost of order $\mathcal{O}(D)$; the large fluctuation above requires $\sim D/2$ beads to shift, so that the statistical cost is of order $\sim \exp(-D^2)$ \cite{Marsh:2011aa}, in agreement with \cite{Aazami:2005jf,Dean:2006,Dean:2008}.     

Applied to inflationary cosmology, we can deduce that inflation is most likely driven by fields near a saddle point, not by fields trapped in a metastable vacuum, or even more unlikely, via chain inflation models \cite{Freese:2004vs,Freese:2006fk} (see \cite{Cline:2011fi} for a recent (critical) feasibility study). This expectation was confirmed in the KKLMMT setup investigated in \cite{Agarwal:2011wm}, as well as for the random potentials of \cite{Frazer:2011tg,Schulz:2011,Frazer:2011br}. 

Another important application is the search for deSitter vacua in string theory that are compatible with a small cosmological constant: uplifted AdS-vacua in type IIA string theory acquire off-diagonal terms in the mass matrix, often pushing the lowest eigenvalue to negative values and thus destabilizing the vacuum \cite{Chen:2011ac}.  

However, one crucial feature is missing in the above arguments: effective potentials used for inflation, as the ones in \cite{Agarwal:2011wm,Frazer:2011tg,Frazer:2011br}, are only valid at low energy  $|V|\ll M_P^4$; thus, by construction, the potentials have to be bounded from above and below to remain in the validity range of effective-field-theory; as a consequence, minima are more likely close to the lower boundary and maxima close to the upper one. The inclusion of this feature is the main goal of this paper, see Sec.~\ref{sec:softboundpot}; we comment on consequences for inflation in Sec.~\ref{sec:inflation}. 

\subsection{Likelihood of connecting two Saddle-Points during Inflation}

 If two or more periods of inflation are to be connected during the last sixty efolds of inflation, as envisioned in folded inflation \cite{Easther:2004ir}, it is necessary that the end of an inflationary trajectory connects to another saddle-point within a distance $H$ inside field-space.  The number of extrema present in the cubic toy-landscape $(2\Delta)^2$ with $\Delta\sim M_p$ introduced in Sec.\ref{sec:review1} falls below one if $\alpha\leq 2D^{1/4}$, so that the separation $r$ between adjacent minima is too large. For larger $\alpha$ the likelihood of finding two adjacent inflationary saddle-points can be estimated as \cite{Aazami:2005jf}
\begin{equation}
P(r^2<H^2)=\frac{1}{\sqrt\pi D}\left[\frac{e}{2D}\frac{1}{D^2}\right]^{D/2}\,,
\end{equation}
under the assumption that the energy scale on field-space is reduced compared to the Planck scale by $1/\sqrt{D}$, $H\sim M^2/M_P\sim 1/D$. This probability decreases sharply as the dimensionality of field space increases. Thus, Azami and Easther concluded that inflation is more likely to happen near a single saddle-point, if at all. Again, this is in line with the recent case studies in \cite{Agarwal:2011wm,Frazer:2011tg,Frazer:2011br}.

\subsection{Likelihood of obtaining N e-folds of Inflation near a Saddle-Point \label{sec:saddle}}
We saw in the last section that inflation on a random potential most likely occurs at or near a single saddle-point. Since, by construction, saddle-points lie within shallow regions, the inflationary trajectory can often be approximated by a straight line in field space (an effective single-field model can be used), traversing an inflection point.

Let us therefore follow \cite{Freivogel:2005vv,Agarwal:2011wm}, where inflation driven by a single field occurs near an inflection point, located at a position where $V''=0$ and $V'$ is small. The origin of $\varphi$ is chosen to correspond to the zero of $V''$, such that
\begin{equation}
V(\varphi)\approx c_0+c_1\varphi+c_3\varphi^3+..., 
\end{equation}
and the $c_i$ are constants.  In the regime where the constant term dominates and the $c_i$ are small, the number of e-folds of inflation is of order
\begin{equation}
N_e\sim \frac{c_0}{\sqrt{c_1c_3}}\times{\cal O}(1)\,.
\end{equation}
To obtain the probability of $N_e$ e-folds of inflation, Freivogel et.al. \cite{Freivogel:2005vv} suggest to compute 
\begin{equation}
P(N_e)=\int\Pi_{i=1}^{\kappa}d\xi_iF(\xi_1,...\xi_{\kappa})\delta\big (N_e-f(\xi_1,...\xi_{\kappa})\big)\,, \label{prob}
\end{equation}
where $\xi_i$ are the model's parameters, $f$ is the number of e-folds as a function of $\xi_i$, and $F$ is  a measure of the parameter space.  Taking a flat measure, $F=(c_0,c_1,c_3)\equiv 1$,  Eq. (\ref{prob}) simplifies to
\begin{equation}
P(N_e)=\int dc_1dc_3\delta\Big(N_e-\frac{c_0}{\sqrt{c_1c_3}}\Big)\,.
\end{equation}
To evaluate the integral, one uses $\int f(x)\delta(g(x))=\int\sum_{i}f(x)\delta(x-x_i)/|g'(x_i)|$, where the sum is over the simple roots of $g(x)$, resulting in
\begin{eqnarray}
P(N_e)=\frac{2c_0^2}{N_e^3}\int \frac{1}{c_3}d\,c_3\,.
\end{eqnarray}
Taking  $-c_0\lesssim c_3 \lesssim c_0$ as the integration boundaries yields the estimate \cite{Freivogel:2005vv,Agarwal:2011wm}
\begin{equation}
P(N_e)\sim -\frac{4c_0^2}{N_e^3}\log(c_0)\,,
\end{equation}
where $c_0$ is small. The crucial feature is 
\begin{equation}
P(N_e)\propto N_{e}^{-3}\,,
\end{equation}
indicating that shorter bursts of inflation are exceedingly more likely than longer ones \footnote{If inflation does not take place near a saddle, but a shallow linear slope, the scaling becomes $P(N_e)\propto N_{e}^{-4}$ \cite{Freivogel:2005vv}. It should be noted that this result depends on the parametrization of the linear slope in \cite{Freivogel:2005vv} and the choice of a flat measure for those parameters. This freedom of choice in the measure is unavoidable in the phenomenological bottom up approach we employ in this study.}. This general feature has been observed in the two dimensional string landscape of \cite{Agarwal:2011wm} as well as in \cite{Frazer:2011tg,Schulz:2011,Frazer:2011br} and Sec.~\ref{sec:inflation}, using the toy landscape of Sec.~\ref{sec:softboundpot}. We would like to emphasise that both, the derivation and the result above, are not new and can be found in \cite{Freivogel:2005vv,Agarwal:2011wm}. 

Such a suppression raises an important question: why does our universe require around $60$ e-folds of inflation instead of say $40$ (which would be more likely) to be consistent with observations? Based on the requirement that gravitationally bound structures form, an anthropic bound on (negative) curvature, and thus the minimum amount of inflation needed, was derived in \cite{Freivogel:2005vv}, yielding $N_{structure} > 59.5$ (see also \cite{Vilenkin:1996ar,Garriga:1998px} for earlier work), close to the observed value.

\section{A Softly Bounded Potential \label{sec:softboundpot}}
From the brief review in Sec.~\ref{sec:review} we learned that saddle-points are more likely than minima or maxima in a random potential (in the absence of upper/lower boundaries), the more so the higher the dimensionality of field space $D$. Further, for $D\gg 1$, inflation is most likely to take place near a single saddle-point (if at all) and last for the shortest duration compatible with observational constraints. 

Here, we would like to investigate how bounds on the potential affect the ratio of saddle-points to extrema. Let us start with the simple potential proposed in \cite{Tegmark:2004qd,Frazer:2011tg,Schulz:2011,Frazer:2011br}
\begin{eqnarray}
 V&=&\sum_{J_1,\dots,J_D=1}^{n}\bigg(a_{J_1\dots J_D}\cos\sum_{i=1}^DJ_i x_i
+b_{J_1\dots J_D}\sin\sum_{i=1}^DJ_i x_i\bigg)\\
&\equiv& \sum_{J_1,\dots,J_D=1}^{n} V_{J_1,\dots,J_D}\,,\label{potential}
\end{eqnarray} 
where $x_i=\varphi_i/M_p$, $M_p=(8 \pi G)^{-1/2}\equiv 1$ and $a_{J_1\dots J_D},b_{J_1\dots J_D}$ are Gaussian random variables
\begin{eqnarray}
P_{a_{J_1\dots J_D}}=\frac{1}{\sqrt{2\pi}\sigma_{J_1\dots J_D}}\exp\left(-\frac{a_{J_1\dots J_D}^2}{2\sigma_{J_1\dots J_D}^2}\right)\label{aijdistribution}
\end{eqnarray}
with zero mean and variance
\begin{eqnarray}
\sigma_{J_1\dots J_D}=\exp\left(-\sum_{i=1}^D\frac{J_i^2}{Dn}\right)\,, \label{variance1}
\end{eqnarray}
and we defined
\begin{eqnarray}
 V_{J_1,\dots,J_D}\equiv a_{J_1\dots J_D}\cos\sum_{i=1}^DJ_i x_i
+b_{J_1\dots J_D}\sin\sum_{i=1}^DJ_i x_i\,.\label{VJ}
\end{eqnarray} 

In essence, the potential is just a truncated Fourier series with suppressed coefficients for higher wavenumbers $J_i\gg 1$, such that inflation can be accommodated \cite{Frazer:2011tg,Schulz:2011,Frazer:2011br}. $J_1=\dots=J_D=0$ is omitted to keep the potential levelled around $V=0$. Opposite to a full Fourier series, only positive integer values for $J_i$ are considered in (\ref{potential}) to keep the number of terms manageable as $D$ increases; this comes at the price of sacrificing spherical symmetry, but will have no qualitative effect onto the results of this paper\footnote{All analytic estimates in this paper could easily be done with the full Fourier series, but numerical comparisons would become much more challenging.}. Further, to ease comparison with \cite{Frazer:2011tg,Frazer:2011br}, we also set all coefficients to zero if a single index vanishes, $J_i=0$ (this choice is not physically motivated, but does not affect our conclusions either). Below the periodicity scale but above $\Delta x \sim 1/\sqrt{Dn}$ \footnote{Coefficients of modes with wavelength below $1/\sqrt{Dn}$ are strongly suppressed, see (\ref{variance1}); thus, the potential is not random on small scales.}, such a potential is certainly random, reminiscent of a beginners sky-slope, i.e. without cliffs or sharp bumps. To ease notation, we use 
\begin{eqnarray}
\mathbf{J}&\equiv& J_1,\dots,J_D\,,\\
\mathbf{x}&\equiv& x_1,\dots, x_D\,, \,\,\, \mbox{etc.}\,,
\end{eqnarray}
in the following.

If points $\mathbf{x}_a$ and $\mathbf{x}_b$ in field space are separated by more than $\Delta x=|\mathbf{x}_{a}-\mathbf{x}_{b}|\sim 1/\sqrt{Dn}$, the value of the potential at such distinct points is again a Gaussian random variable with zero mean. The easiest way to see this is to consider first the same point $\mathbf{x}_c$ in different realizations of the random potential (i.e.~consider an ensemble): the $V_{\mathbf{J}}$ are the sum of two random, Gaussian variables, $a_{\mathbf{J}}$ and $b_{\mathbf{J}}$, with zero mean, variance according to (\ref{variance1}) and constant coefficients (the cosine and sine factors). As a consequence, the $V_{\mathbf{J}}$ are also Gaussian random variables with zero mean and the same variance
\begin{eqnarray}
\sigma_{V_\mathbf{J}}&=&
\sqrt{\sigma_\mathbf{J}^2
\left(\left(\frac{\partial V_\mathbf{J}}{\partial a_\mathbf{J}}\right)^2
+\left(\frac{\partial V_\mathbf{J}}{\partial b_\mathbf{J}}\right)^2\right)}
=\sigma_\mathbf{J}\,,
\end{eqnarray}
where the location $\mathbf{x}_c$ dropped out since 
\begin{eqnarray}
\left(\frac{\partial V_\mathbf{J}}{\partial a_\mathbf{J}}\right)^2
+\left(\frac{\partial V_\mathbf{J}}{\partial b_\mathbf{J}}\right)^2=\left(\cos\sum_{i=1}^DJ_i x_i^c\right)^2+\left(\sin\sum_{i=1}^DJ_i x_i^c\right)^2=1\,.
\end{eqnarray}
Therefore, the value of the potential at a given point
\begin{eqnarray}
V(\mathbf{x}_c)=\sum_{\mathbf{J}}V_\mathbf{J}(\mathbf{x}_c)
\end{eqnarray}
is also a Gaussian random variable with zero mean and variance
\begin{eqnarray}
\sigma_V^2=\sum_\mathbf{J}\sigma_\mathbf{J}^2\approx \left(\frac{Dn\pi}{8}\right)^{D/2}\,. \label{varianceV}
\end{eqnarray}
To evaluate the sum, we first plugged in (\ref{variance1}), approximated the sum by an integral which we could then evaluated in the limits
$(n+1)\sqrt{2/(Dn)}\gg 1$ and $\sqrt{2/(Dn)}\ll 1$, that is for
\begin{eqnarray}
 n\gg D\,.
 \end{eqnarray}
 Let us now turn our attention to \emph{distinct locations in the same random potential}: as long as points are separated more than $1/\sqrt{Dn}$ (but less than the periodicity scale), the corresponding values of the potential become uncorrelated; however, since the coefficients $a_{\mathbf J}$ and $b_\mathbf{J}$ are chosen only once, the potential is strictly bounded from above and below (sines and cosines are bounded); nevertheless, for $|V|\ll \sum_{\mathbf{J}}\sigma_{\mathbf J}\approx (Dn\pi/4)^{D/2}$ we can approximate the probability distribution of $V$ by a Gaussian in the limit of large $n$ and $D$ (invoking the central limit theorem); we may further approximate the variance of this Gaussian by (\ref{varianceV}) obtained from \emph{different realizations of the potential at the same point}, that is obtained by considering an ensemble of potentials\footnote{Note that $\sum_{\mathbf{J}}\sigma_{\mathbf J}/\sigma_{V}\approx (Dn\pi/2)^{D/4}\gg 1$, providing a wide validity range. No important results in this paper depend on the actual value of $\sigma_V$ in (\ref{varianceV}), and we treat $\sigma_V$ as a free parameter in later sections.}. 
 
 This reasoning based on an ensemble is identical to the one used in \cite{Frazer:2011tg,Frazer:2011br}: to investigate inflation, the same starting point $\mathbf{x}=0$ was used in different realizations of the random potential, instead of sampling different initial conditions in a single realization of the potential. The use of ensembles is of course nothing new and lies, for instance, at the heart of statistical mechanics. We shall use an ensemble of potentials again in Sec.~\ref{sec:hessian} to compute properties of the Hessian at critical points.
 
In the limit of large $Dn$, $1/\sqrt{Dn}$ approaches zero, indicating that the value of the potential at any point is an independent random variable. Obviously, such a potential would not allow slow roll inflation to take place and is therefore of little interest to us. Thus, we use $n=5$ for all numerical computations in this paper, while treating the value of the potential at distinct critical points as independent, Gaussian random variables with zero mean and variance according to (\ref{varianceV}). By distinct we mean that critical points are separated more than $1/\sqrt{Dn}$; if they happen to be closer, we are essentially dealing with the same flat region in field space, rendering a distinction physically meaningless.

An important property of our potential is that it is well approximated by a softly bounded one: we expect saddle-points to be ubiquitous for $|V|\ll \sigma_V$, in line with the conclusions of Sec.~\ref{sec:review1}, whereas minima/maxima should become more likely for  $|V|\gg \sigma_V$, which we shall prove now.

\subsection{The Hessian \label{sec:hessian}}
We denote the value of the potential in (\ref{potential}) at a critical point $\bar{\mathbf{x}}$ by
 $V(\bar{\mathbf{x}})\equiv \bar{V}$, that is,
\begin{eqnarray}
\bar{V}=V_{1,\dots,1}(\bar{\mathbf{x}})+\sum_{\substack{\mathbf{J}=1
,\\ \lnot J_1=\dots=J_D=1 }}^n V_\mathbf{J}(\bar{\mathbf{x}})\,,
\end{eqnarray} 
 which, according to (\ref{varianceV}), is a Gaussian variable with variance $\sigma_V^2\approx \left(Dn\pi/ 8\right)^{D/2}$.
 We need to enforce this relationship if we want to retain the boundedness of the potential; to this end, we chose to replace $V_{1,\dots,1}(\bar{\mathbf{x}})$ by $\bar{V}$ and the shortened sum above. Hence, the Hessian becomes
\begin{eqnarray}
H_{kl}&\equiv&\frac{\partial^2V}{\partial_k \partial_l}\\
&=&-\bar{V}-\sum_{\mathbf{J}=1}^n\left(J_kJ_l-1\right)V_J\\ \label{fullHessian}
&\equiv& -\left(\bar{V}+h_{kl}\right)\,,
\end{eqnarray} 
for $k,l=1,\dots,D$; the sum covers the full range again, since $J_1=\dots=J_D=1$ does not contribute. The $h_{kl}$ are Gaussian distributed with zero mean and variance
\begin{eqnarray}
\sigma_{kl}^2\approx \Bigg\{ \begin{array}{l}
\left(\frac{Dn\pi}{8}\right)^{D/2}\frac{3}{16}D^2n^2\equiv \sigma^2\,\, \mbox{for}\,\, k=l \\
\\
\sigma^2/3\,\, \mbox{for}\,\, k\neq l\,,
\end{array}
\end{eqnarray} 
where we replaced sums by integrals, which we evaluated again in the limit $n\gg D$. It should be noted that even though the $h_{kl}$ are not independent, we will treat them as such in the following, based on universality in random matrix theory.\footnote{Correlations between the $h_{kl}$ arise due to the limited number ($2(n^D-D)$) of terms in the potential, as well as the constraints $\nabla V|_{\mathbf{x}=\bar{\mathbf{x}}}=0$, which we do not impose.  This harsh approximation becomes good if the potential contains a large number of independent Gaussian parameters, that is in the limit $n\gg D$.} 

Thus, given a critical point $\bar{\mathbf{x}}$ with $V(\bar{\mathbf{x}})=\bar{V}$, the Hessian is a symmetric matrix of Gaussian distributed real numbers with mean $-\bar{V}$, variance $\sigma$ on the diagonal and $\sigma/\sqrt{3}$ otherwise. The novel feature is the presence of a non-zero mean for matrix entries which encodes directly the presence of the soft upper and lower boundaries in the potential, which entered by enforcing the constraint $V(\bar{\mathbf{x}})=\bar{V}$.

\begin{figure}[tb]
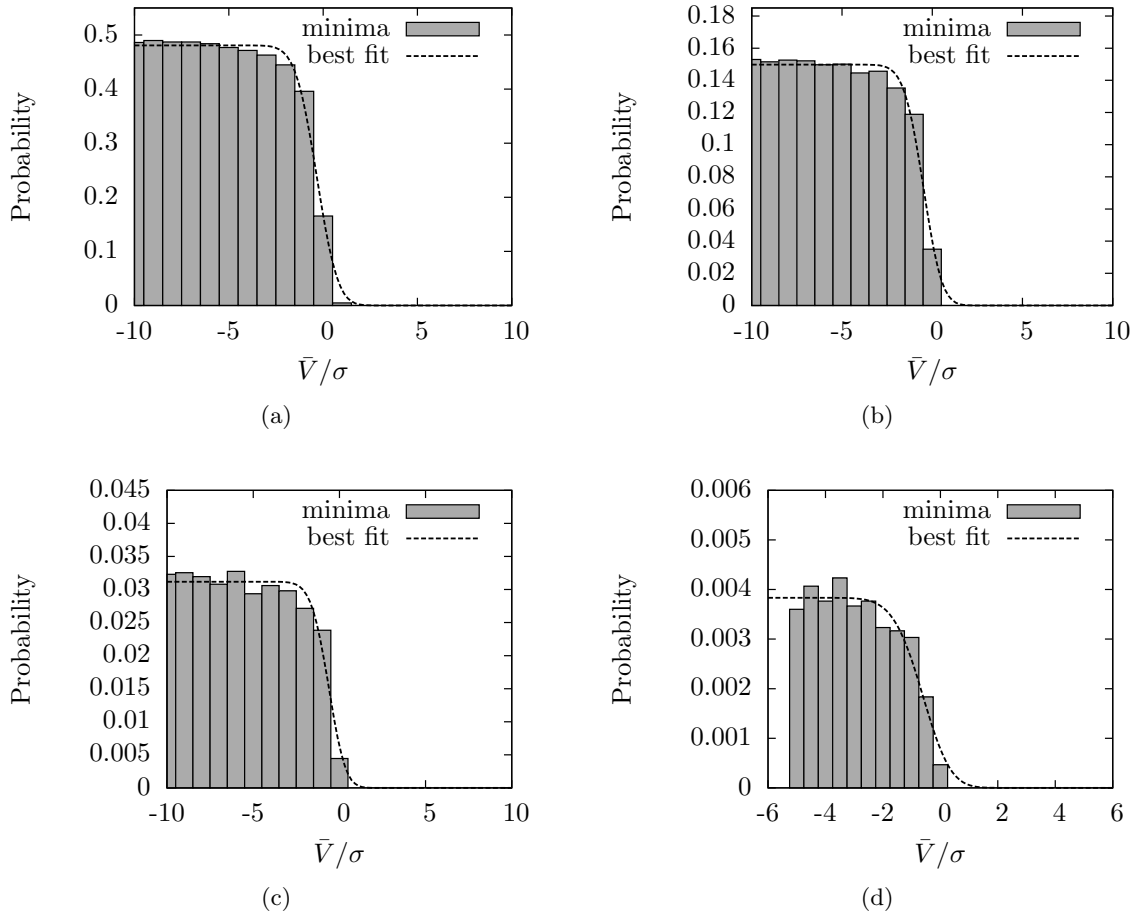

 \begin{tabular}{cc}
 
\subfloat[]
{\label{fig:matrix2a}
\scalebox{0.92}{\input{matrix2a}}} &

\subfloat[]{
\label{fig:matrix3a}
\scalebox{0.92}{\input{matrix3a}}} \\

\subfloat[]{
\label{fig:matrix4a}
\scalebox{0.92}{\input{matrix4a}}} &

\subfloat[]{
\label{fig:matrix5a}
\scalebox{0.92}{\input{matrix5b}}} 

\end{tabular}
\caption{The probability $P_{min}$ of a positive definite Hessian (a critical point is a minimum) is computed for an ensemble of random matrices with $H_{ij}=-\bar{V}-h_{ij}$ for (a) $D=2$ and $10^5$ matrices per bin,  (b) $D=3$ and $5\cdot 10^4$ matrices per bin, (c) $D=4$ and $10^4$ matrices per bin, (d) $D=5$ and $3\cdot 10^4$ matrices per bin; $h_{ij}=h_{ji}$ are Gaussian random variables with zero mean, variances $\sigma_{ii}=\sigma$ on the diagonal and $\sigma_{ij}=\sigma/\sqrt{3}$ for $i\neq j$, to mimic the Hessian of our toy random potential in (\ref{potential}). The dotted line is a two-parameter fit based on the Ansatz in (\ref{conjecturedprobability}), resulting in $c_{\mbox{\tiny num}}$ and $p_{\mbox{\tiny num}}$ in Tab.~\ref{tab:Pmin}, where we also compare $P_{min}(-\infty)$ to the upper bound in (\ref{Pminlimit}) with (\ref{cana}) and $p=0.6$. The $D=5$ case appears noisy due to $P_{min}\sim 10^{-3}$ while drawing only $3\cdot 10^4$ matrices per bin, see Fig.~\ref{fig:matrix5err}. \label{fig:matrices2to5}}
\end{figure}

Since we are primarily interested in the effect of $\bar{V}$, let us consider an even simpler Hessian by ignoring the off diagonal elements when computing eigenvalues, which become
\begin{eqnarray}
\lambda_k  = -(\bar{V}+h_{kk}) \,.
\end{eqnarray} 
Consequently, the probability that a critical point at height $\bar{V}$ is a minimum becomes
\begin{eqnarray}
P_{min}=\left(\frac{1}{2}\left(1-\mbox{erf}\left(\bar{V}/(\sqrt{2}\sigma)\right)\right)\right)^D\,;
\end{eqnarray}
the probability for a maximum follows by replacing $\bar{V}$ by $-\bar{V}$, and the chance for a saddle-point is $P_{sad}=1-P_{min}-P_{max}$. For $\bar{V}=0$, we recover the result of (\ref{psaddle1}), $P=(1/2)^D$, that is saddle-points are the most likely occurrence. However, for large negative values of the potential, $\bar{V}\ll \sigma\sqrt{2}\, \mbox{erf}^{-1}(1-2^{1-1/D})\approx -\mbox{few} \times \sigma$  (i.e.~for $D\sim 10^3$), the probability that a critical point is a minimum approaches one.

For the particular potential in (\ref{potential}) the variance scales as $\sigma \propto D^{D/4}Dn$, so the range of ubiquitous 
saddle-points increases rapidly with increasing $D$, but if a different scaling of the coefficients' variance $\sigma_\mathbf{J}$ were chosen, one could alter this scaling easily. To keep our conclusions in subsequent sections as general as possible, we treat $\sigma$ as an unspecified free parameter from here on.

\begin{figure}[tb]

\begin{center}
\scalebox{0.92}{\input{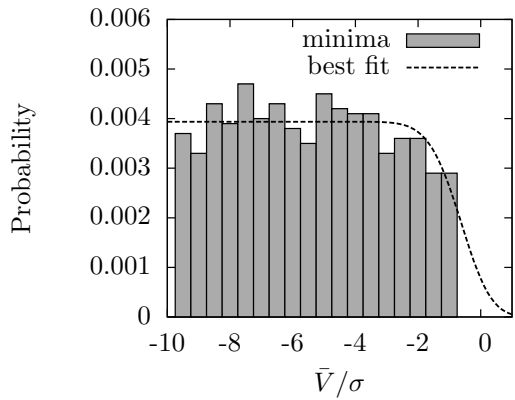}} 
\end{center}

\caption{The probability $P_{min}$ as in Fig.~\ref{fig:matrices2to5} (d) for $D=5$ and smaller values of $\bar{V}$ (a new drawing of random matrices is used); we observe that any features below $\bar{V}/\sigma \lesssim -4$ are indeed only due to noise (as expected), caused by drawing only $10^{4}$ matrices per bin. \label{fig:matrix5err}}
\end{figure}

Before we discuss consequences for inflation, lets step back and consider the more general, non-diagonal Hessian in (\ref{fullHessian}); in this case, one could go the arduous route of computing the joint probability distribution (JPD) of eigenvalues and investigate the probability of a large fluctuation leading to a positive definite Hessian, as in \cite{Marsh:2011aa}. However, a simple Ansatz (tested numerically) suffices for our purposes:  as we saw in Sec.~\ref{sec:review1}, the probability at $\bar{V}=0$ is quite universally of the form $P\propto \exp(-cD^p)$ with $c=\mathcal{O}(1)$ and $1<p\lesssim 2$ for JPD not too different from the Gaussian unitary ensemble. In the limit $|\bar{V}|\ll \sigma$, we could use the result of \cite{Chen:2011ac} in (\ref{mixedHessian}) if $y\ll 0.1$, to investigate at which $D$ the Gaussian suppression starts to dominate, i.e.~when $Dc_2/c_1>1$; however, for our particular potential the ratio of the variances is $y=1/\sqrt{3}$, naively indicating that the Gaussian suppression $\propto e^{-D^2}$ dominates for any value of $D$, but the analytic estimate in (\ref{mixedHessian}) is of course not applicable any more. 

To heuristically add the effect of a softly bounded potential, we make the Ansatz
\begin{eqnarray}
P_{min}\sim \left(e^{-c}\left(1-\mbox{erf}\left(\frac{\bar{V}}{\sqrt{2}\sigma}\right)\right)\right)^{D^p}\,,\label{conjecturedprobability}
\end{eqnarray}   
with $p$ and $c$ as free parameters (specific to the Hessian at hand). For $\bar{V}\ll -\sigma$, the probability is bounded from above by the limiting value  $P_{min}^{\mbox{\tiny limit}}\sim (2e^{-c})^{D^p}$. For $D=1,2$ one can easily compute this limit analytically, leading to $c_{D=1}=\ln(2)$ and $c_{D=2}=\ln(2)(1+1/2^p)$. We tested (\ref{conjecturedprobability}) numerically for $D=2,\dots,5$, see Fig.~\ref{fig:matrices2to5} and Tab.~\ref{tab:Pmin}, finding  good agreement between the shape of our Ansatz and the full numerical result if we choose $p\approx 0.6$. Further, by setting the normalization factor to\footnote{This is motivated by analytically investigating eigenvalues for $D=3$ in the limit $\bar{V}\ll -\sigma$.}
\begin{eqnarray}
c\equiv  \ln(2)\left(1+\frac{D-1}{D^p}\right)\,\label{cana}
\end{eqnarray}
we arrive at an upper bound of encountering a minimum
\begin{eqnarray} 
P_{min}^{\mbox{\tiny limit}}(V\rightarrow -\infty)\equiv 2^{-D+1}\label{Pminlimit}\,,
\end{eqnarray} 
which is compared with the numerical values 
$P_{min}^{\mbox{\tiny num}}(V\rightarrow -\infty)$ in Tab.~\ref{tab:Pmin}. One could improve the fit by including two exponents $p_1$ and $p_2$, taking over some of the $D$-dependence of $c$ at the expense of a second parameter, but we would like to keep the Ansatz as simple as possible (with the above $c$ the exponential suppression is underestimated). Higher values of $D$ become hard to treat numerically, due to the strong suppression of $P_{min}$; hence, we use (\ref{conjecturedprobability}) with the above $c$ for analytic estimates in the next sections, particularly to derive a lower limit on the magnitude of the expected contribution to the cosmological constant after an inflationary trajectory settled in a local minimum.
One should keep in mind that the parameters are model dependent,  but the (super)exponential suppression of $P_{min}(-\infty)$ is generic.

\begin{table}[tb]
\caption{Comparison of the Ansatz in (\ref{conjecturedprobability}) with $c_{\mbox{\tiny limit}}$ in (\ref{cana}) and $p=0.6$ (leading to the upper bound on $P_{min}^{\mbox{\tiny limit}}$ in (\ref{Pminlimit})) to an ensemble of random matrices with $H_{ij}=-\bar{V}-h_{ij}$; $h_{ij}=h_{ji}$ are Gaussian random variables with zero mean, variances $\sigma_{ii}=\sigma$ on the diagonal and $\sigma_{ij}=\sigma/\sqrt{3}$ for $i\neq j$. See also Fig.~\ref{fig:matrices2to5}. As expected, the probability of encountering a minimum is even more suppressed than $P_{min}^{\mbox{\tiny limit}}$ , since the Hessian is far from diagonal.  }
\begin{center}
\begin{tabular}{|c|ll|l|ll|}
\hline
Dimension & $c_{\mbox{\tiny num}}$ & $c_{\mbox{\tiny limit}}$& $p_{\mbox{\tiny num}}$&$P_{min}^{\mbox{\tiny num}}(-\infty)$& $P_{min}^{\mbox{\tiny limit}}(-\infty)$\\
\hline
$1$ & -- & $\ln(2)\approx 0.69$ & -- &  -- &1\\
$2$ & 1.17 & $1.15$ & 0.61 & 0.48 &$0.5$\\
$3$ & 1.69 & $1.41$ & 0.59 & 0.14 &$0.25 $\\
$4$ & 2.20 & $1.60$ & 0.60 & 0.033 &$0.125$\\
$5$ & 3.12 & $1.75$ & 0.52 & 0.0038 &$0.0625$\\
$D\gg 1$ & -- & $ (1+D^{1-p}-D^{-p})\ln(2)$ & --& --  &$ 2^{-D+1}$\\
\hline\hline
\end{tabular}
\label{tab:Pmin}
\end{center}
\end{table}

\subsection{Where are the Minima? \label{sec:minima}}
Since minima become exceedingly rare as $D$ is increased, one may wonder at which height of the potential most of them lie. We will first give an upper bound of this height based on the results of Sec.~\ref{sec:hessian}, which we subsequently compare with numerics.

\begin{figure}[tb]
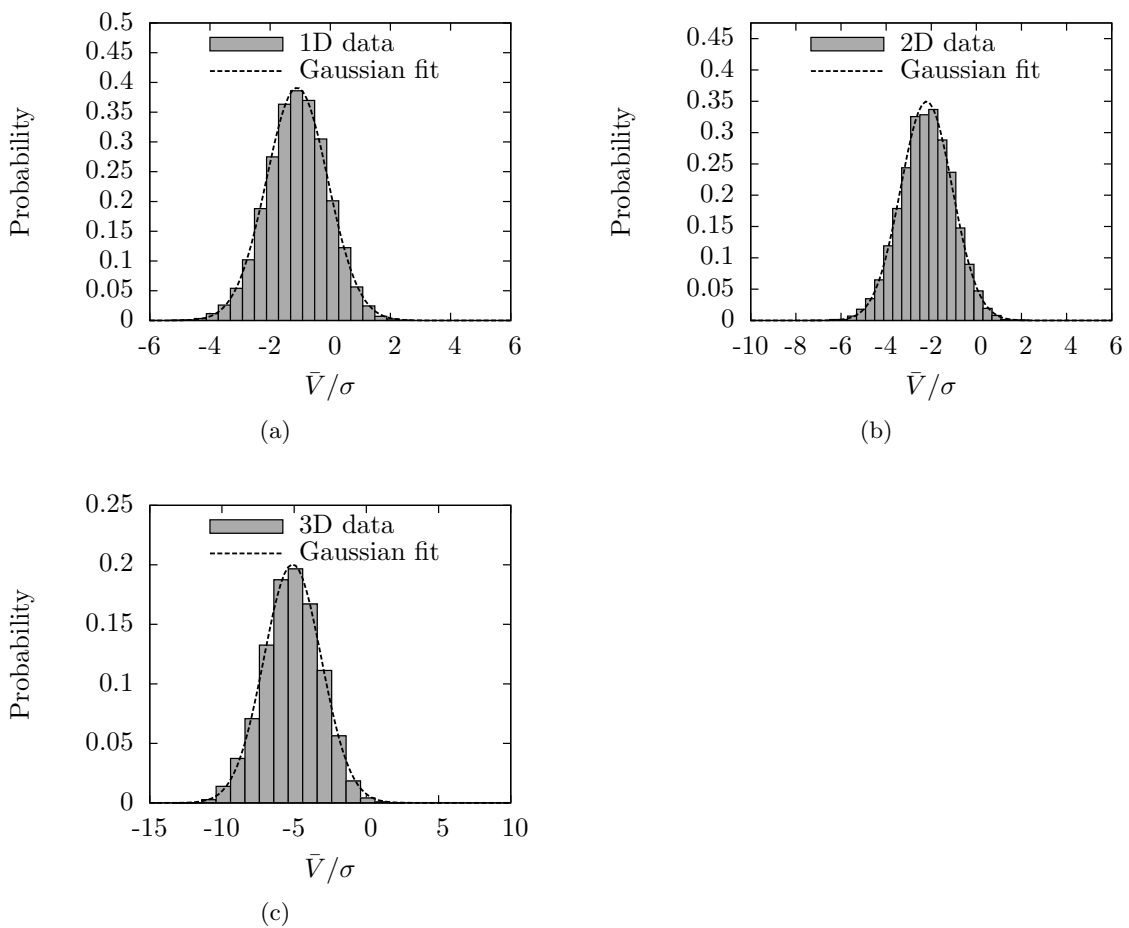

 \begin{tabular}{cc}
 
\subfloat[]
{\label{fig:minima-evolution5a}
\scalebox{0.92}{\input{histogram5a}}} &

\subfloat[]{
\label{fig:minima-evolution6a}
\scalebox{0.92}{\input{histogram6a}}} \\

\subfloat[]{
\label{fig:minima-evolution7a}
\scalebox{0.92}{\input{histogram7a}}} 

\end{tabular}
\caption{The probability of finding a minimum at height $\bar{V}$ is plotted over $\bar{V}/\sigma$ for the potential in (\ref{potential}) and $D=1,2,3$: we generate $10000$ realizations for each $D$ and search for minima according to the algorithm in Sec.~\ref{sec:minima}. The dotted line is a Gaussian fit, whose mean is compared to the analytic upper bound of (\ref{Vp}) in Tab.~\ref{tab:Vp}. \label{fig:minima-evolution}}
\end{figure}

Consider our toy potential in (\ref{potential}), which was chosen to be relatively smooth by suppressing higher frequencies with the aim to aid inflation. What is the number of critical points $n_c$ that are encountered if traversing a potential difference of order $\Delta V\sim \sigma$? Noting $\left<|\nabla V|\right>\sim \sigma$, we deduce $n_c\sim 1$.  We keep $n_c$ general for the time being, keeping in mind that it is a characteristic of the potential \footnote{We consider classes of potentials for which $n_c$ is independent of $D$.}. 

Since the error-function is a bit cumbersome to deal with, we approximate $P_{min}$ in (\ref{conjecturedprobability}) by a step function 
\begin{eqnarray}
P_{min}(\bar{V})\approx 2^{1-D}\theta(\bar{V}_c-\bar{V})\,,
\end{eqnarray} 
where 
\begin{eqnarray}
\bar{V}_c\equiv \sigma\sqrt{2}\,\mbox{erf}^{-1}\left(1-e^{c}\left(2^{-D/D^p}\right)\right)
\end{eqnarray}
and we used $c$ from (\ref{cana}); $\bar{V}_c$ is chosen such that $P_{min}(\bar{V}_c)=P_{min}^{\mbox{\tiny limit}}(-\infty)/2$. The probability of not finding a minimum when moving down the potential  can thus be approximated by
\begin{eqnarray}
P_{no\, min}(\bar{V})\approx (1-P_{min}(-\infty))^{-(\bar{V}-\bar{V}_c) n_c/\sigma}
\end{eqnarray}
for $\bar{V}<\bar{V}_c$; if the distribution of the number of minima versus height is close to a Gaussian, which is the case here, we can estimate the position of the peak $\bar{V}_p$ by setting $P_{no\, min}(\bar{V}_p)=1/2$ and solving for \begin{eqnarray}
\bar{V}_p=\bar{V_c}+\frac{\sigma}{n_c}\frac{\ln(2)}{\ln(1-2^{-D+1})} \label{Vp}\,.
\end{eqnarray}
For large $D$ we get $\bar{V}_p\approx -\sigma 2^{D-1}\ln(2)/n_c$ as the leading order contribution. For $D=10$, this approximation is already better than one percent. Since we used the upper bound for the probability of finding a minimum in (\ref{conjecturedprobability}) with (\ref{cana}) and $p\approx 0.6$, the above provides a conservative upper limit on the height where most minima are encountered and thus a lower bound on the magnitude of the contribution to the cosmological constant.

To check this bound, we generated $10000$ realizations of the potential in (\ref{potential}) for $D=1,2,3$ and $n=5$. To mimic the search of a minimum during inflation, we evolve the field equations in an expanding background\footnote{This method for finding minima is biased towards lower values of the potential; however, this search algorithm yields exactly those minima relevant for the vacuum selection during inflation.}, see Sec.~\ref{sec:inflation}, starting from $\bf{x}=0$; we lift the potential up by $5\sigma$ to keep Hubble friction high. Whenever the field comes to rest, we note the height of the un-shifted potential, and repeat.  

The analytic bound of the peak position is compared to the one in the histograms (Fig.~\ref{fig:minima-evolution}) in Tab.~\ref{tab:Vp}; we observe that $\bar{V}_p$ does indeed provide the desired bound according to the limited range of $D$ that are directly accessible for comparison. We further observe that the variance of the histograms is comparable to the location of the peak,
\begin{eqnarray}
\sigma_{min}=\mathcal{O}(\bar{V}_p)\,, 
\end{eqnarray}
which is expected given the gentle increase of the full $P_{min}$ in (\ref{conjecturedprobability}) from zero to $\bar{V}_p$.  

\begin{table}[tb]
\caption{The peak position of the histograms in Fig.~\ref{fig:minima-evolution} is compared to the analytic upper bound in (\ref{Vp}) with $n_c=1$ and for different dimensions of the field space $D$. The histograms are the result of searching for minima in $10000$ realizations of the potential in (\ref{potential}).}
\begin{center}
\begin{tabular}{|c|cc|}
\hline
Dimension & $\bar{V}_{p}^{\mbox{\tiny num}}/\sigma$ & $\bar{V}_p^{\mbox{\tiny limit}}/\sigma$\\
\hline
$1$ & $-1.1$ & $-1$  \\
$2$ & $-2.2$ & $-1.5$ \\
$3$ & $-5.1$ & $-3.1$ \\
$D\gg 1$ & -- & $-2^{D-1}\ln(2)$ \\
\hline\hline
\end{tabular}
\label{tab:Vp}
\end{center}
\end{table}

As a direct consequence of the exponential suppression of $P_{min}(-\infty)=2^{-D+1}$, the position of the peak $\bar{V}_p$ shifts rapidly to increasingly negative values as $D$ increases, even for our upper bound: for as little as ten fields we already have $\bar{V}_p(10)/\sigma\approx -356$ and for field numbers commonly encountered in the string landscape, $D\sim 500$, we find $\bar{V}_p(500)/\sigma\approx -10^{150}$. Thus, if inflation were to take place in such a potential, the ultimate resting place of the fields would most likely be in contradiction to a positive cosmological constant, even for extremely small $\sigma$:

\vspace{0.3cm}
\begin{center}
 \emph{Opposite to our naive expectation, the likelihood of finding a positive cosmological constant decreases rapidly as $D$ increases -- the huge growth of the absolute number of critical points is irrelevant. } 
\end{center}
\vspace{0.3cm}
This is one of the main results of this paper which we test numerically in Sec.~\ref{sec:inflation} for low $D$. We comment further on the consequences for vacuum selection and the feasibility of multi-field inflation in general in Sec.\ref{sec:cc}. 

We would like to stress again that we severely overestimated the probability of finding a minimum; as a consequence, the above problem of encountering a large negative contribution to the cosmological constant is more severe in a realistic landscape.

\section{Inflation in a Random Potential\label{sec:inflation}}

We generate ensembles of random potentials defined in (\ref{potential}) with $n=5$ and $D=1,2,3$ to investigate the feasibility of inflation and its expected duration. The equations of motion are the Friedmann equations in a homogeneous and isotropic universe ($k=0$),
\begin{eqnarray}
3H^2=V+\sum_{i=1}^{D}\frac{\dot{\varphi}_i^2}{2}\,,\label{Friedmann}
\end{eqnarray}
and the Klein-Gordon equations in an expanding universe
\begin{eqnarray}
\ddot{\varphi_i}+3H\dot{\varphi_i}=-\frac{\partial V}{\partial\varphi_i}\,.\label{KleinGordon}
\end{eqnarray}
We are interest in the amount of inflation as measured by the number of e-foldings
\begin{eqnarray}
N=\int H(t)\,dt\,.\label{efolds}
\end{eqnarray} 

\subsection{A Cut-off for large $D$ \label{sec:cutoff}}
Since the number of terms in the potential increases fast even for our already truncated Fourier series in (\ref{potential}),
\begin{eqnarray}
N_{\mbox{\tiny Pot}}\equiv 2(n^D-D)\,,
\end{eqnarray}
 performing many runs becomes computationally challenging. To ameliorate this technical problem, one can truncate the series by setting coefficient $a_{\mathbf{J}}$ and $b_{\mathbf{J}}$ to zero if their magnitude, chosen randomly according to (\ref{aijdistribution}) with variance (\ref{variance1}), is below a prescribed cut-off. As a consequence, the potential becomes smother; hence, one should be careful to choose the cut-off not too high in order to preserve the qualitative features of inflationary regimes. Practically, we would like to keep a small fraction of coefficients only, $F<1$, which we can identify on geometric grounds with
\begin{eqnarray}
F= \frac{V_DJ_{max}^D}{2(n^D-D)}\,,
\end{eqnarray} 
where 
\begin{eqnarray}
J\equiv |\mathbf{J}|=\sqrt{\sum_{i=1}^DJ_i^2}\,,
\end{eqnarray}
$V_D=\pi^{D/2}/\Gamma(D/2+1)$ is the volume of the unit sphere in $D$ dimensions, $\Gamma$ is the Gamma function ($\Gamma(x)\approx \sqrt{2\pi}x^{x-1/2}e^{-x}$ if it is expanded for large arguments) and we assumed $J_{max}<n$ \footnote{For $J_{max}>n$, the sphere does not fit into the cube with diameter $2n$ and we would count coefficients already set to zero in the potential (\ref{potential}).}. Since the variance in (\ref{variance1}) is a monotonically increasing function of $J$, we expect that most coefficients with $J>J_{max}$ will have values smaller than
\begin{eqnarray}
\sigma_{min}&\equiv &\sigma_{{\mathbf J}_{max}}\\
&=&\exp\left(\frac{-J_{max}^2}{Dn}\right)\\
&=&\exp\left(-4\left(\frac{F}{\pi^D/2}\Gamma(D/2+1)\right)^{2/D}\right)\,.
\end{eqnarray}

\begin{figure}[tb]
 \begin{tabular}{c}
 
\subfloat[]
{\label{fig:cutoffcomparison1}
\scalebox{1}{\input{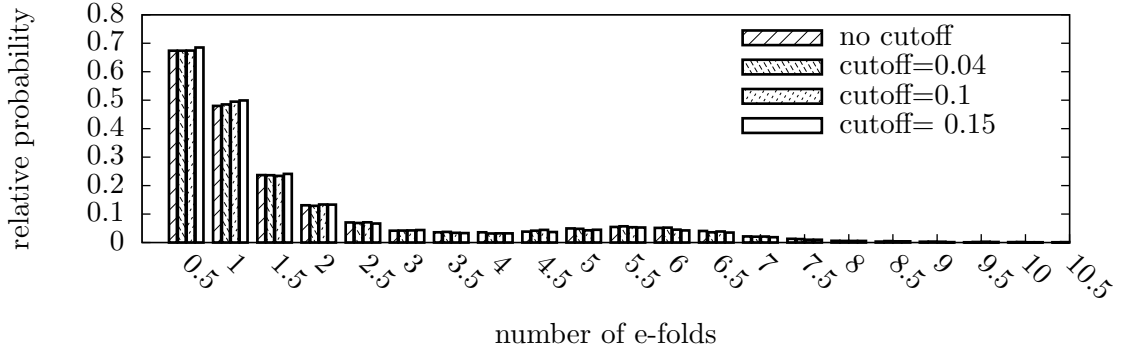}}} \\

\subfloat[]{
\label{fig:cutoffcomparison2}
\scalebox{1}{\input{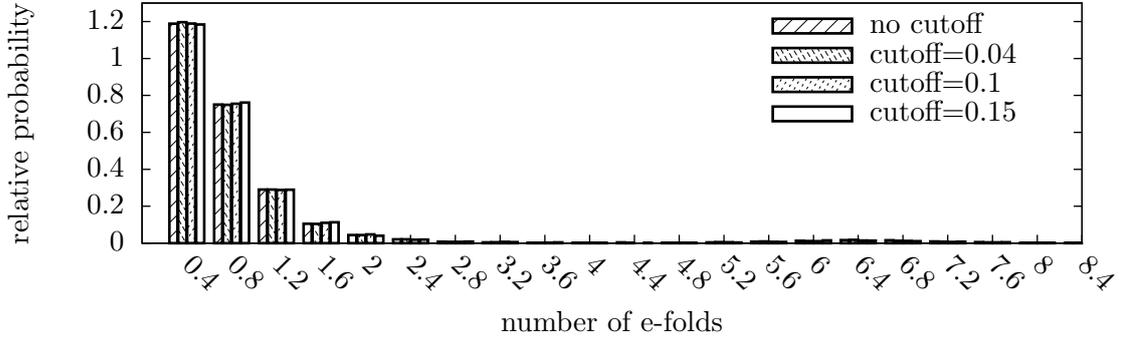}}} \\

\subfloat[]{
\label{fig:cutoffcomparison3}
\scalebox{1}{\input{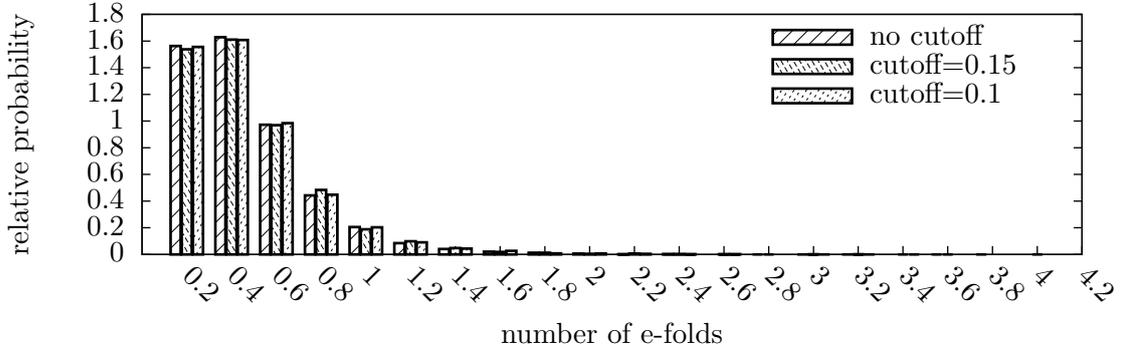}}} 

\end{tabular}
\caption{The probability of $N$-efolds of inflation in random potentials (\ref{potential}) with $n=5$ is computed with and without imposing a cut-off $\sigma_{min}=0,0.04,0.1,0.15$ onto the coefficients $a_{\mathbf J}$ and $b_{\mathbf J}$ for $D=1,2,3$ (top to bottom), see Sec.~\ref{sec:cutoff}. We find no statistical significant effect for $\sigma_{min}\lesssim 0.15$, justifying the use of a cut-off.\label{fig:cutoffcomparison}}
\end{figure}

How big can one choose $\sigma_{min}$ without drastically altering the evolution of fields? The energy scale in our potential is set by the coefficients with the lowest $J$ and thus of order $V\sim 1$. If we want to discuss inflation compatible with observations, we would need to leave the dynamics untouched even if the kinetic energy $K=\sum_i\dot{\varphi_i}^2/2$ is only a small fraction of $V$, i.e.~slow roll suppressed $K\sim \epsilon V \sim 10^{-2}$.  Coefficients $a_{\mathbf J}, b_{\mathbf J}\lesssim K\sim 10^{-2}$ will at most lead to a hill/valley that is easily traversed, even on an inflationary slow roll trajectory. Thus, imposing a cut-off of order
\begin{eqnarray}
\sigma_{min} \lesssim 10^{-2} 
\end{eqnarray} 
should leave conclusions with regard to inflationary dynamics unchanged. 

We tested numerically the effect of $\sigma_{min}=0,0.04,0.1,0.15$ for $D=1,2,3$ onto the relative probability of finding $N$ e-folds of inflation, see Fig.~\ref{fig:cutoffcomparison}, and found no statistically significant effect for cut-offs as large as $\sigma_{min}=0.15$. An explanation might be that more expansion requires a flatter potential, that is the absence of high frequency oscillations. Hence, all runs with sizable amount of inflation operate on potentials with $a_{\mathbf J}$ and $b_{\mathbf J}$ close to zero for larger $J$; these coefficients are thus unaffected by increasing the cut-off to relatively large values.  

\subsection{The Feasibility of Inflation \label{sec:feasibility}}
\begin{figure}[tb]
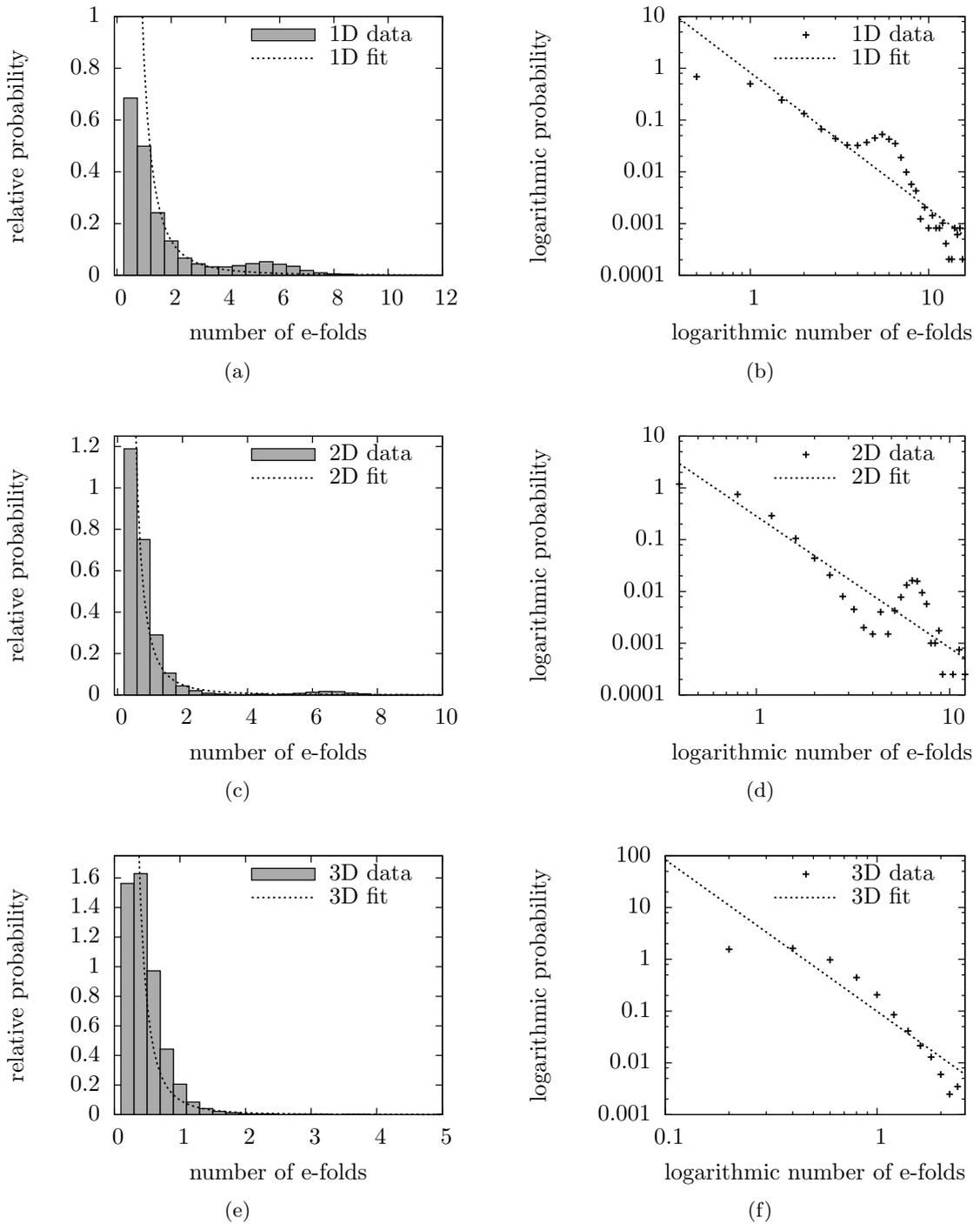

 \begin{tabular}{cc}
 
\subfloat[]
{\label{fig:cutoffcomparisona}
\scalebox{0.92}{\input{fit-test1a}}} &

\subfloat[]
{\label{fig:cutoffcomparisonb}
\scalebox{0.92}{\input{fit-test1}}} \\

\subfloat[]
{\label{fig:cutoffcomparisons}
\scalebox{0.92}{\input{fit-test2a}}} &

\subfloat[]
{\label{fig:cutoffcomparisond}
\scalebox{0.92}{\input{fit-test2}}} \\

\subfloat[]
{\label{fig:cutoffcomparisone}
\scalebox{0.92}{\input{fit-test3a}}} &

\subfloat[]
{\label{fig:cutoffcomparisonf}
\scalebox{0.92}{\input{fit-test3}}} 

\end{tabular}
\caption{The probability of $N$-efolds of inflation in  random potentials (\ref{potential}) with $n=5$ is compared to a power-law $P(N)=\alpha N^{-\beta}$ with coefficients in Tab.~\ref{sec:saddle} for $D=1,2,3$, according to the algorithm in Sec.~\ref{sec:feasibility}. For each $D$ we perform $20000$ runs. The relative importance of the bump like feature diminishes as $D$ is increased. \label{fig:PNtest}} 
\end{figure}

Based on inflation near a saddle-point, we saw in Sec.~\ref{sec:saddle} that the probability of inflation should decrease sharply with the number of e-folds, $P\propto N^{-3}$, independent of the dimensionality $D$ of field-space. This behaviour can indeed be observed qualitatively in our toy potential (\ref{potential}); the algorithm we use for counting the number of e-folds for a given $D$ is
\begin{enumerate}
\item Choose the coefficients $a_{\mathbf{J}},b_{\mathbf{J}}$ according to (\ref{aijdistribution}) with variance (\ref{variance1}). If $D\geq 2$, impose a cut-off $\sigma_{min}\equiv 0.14$ and set all $a_{\mathbf{J}},b_{\mathbf{J}}$ to zero if their magnitude is below $\sigma_{min}$.  
\item Choose the initial field values and velocities\footnote{Choosing the same starting point is justified since each run a new potential is generated.} as $\varphi_i=\dot{\varphi}_i=0$ for $i=1,\dots,D$. If $V(0)<0$ reject the potential and go back to step 1.
\item Evolve the equations of motion (\ref{Friedmann}) and (\ref{KleinGordon}) numerically and compute the number of e-folds in (\ref{efolds}).
\item Stop the evolution if $V<0$ or the fields get stuck in a minimum (i.e.~no discernible movement within one e-fold). Save the number of e-folds up to this time and go back to step 1.
\end{enumerate} 
We plot the resulting probability for a large number of runs in Fig.~\ref{fig:PNtest} together with the best fit $P=\alpha N^{-\beta}$ yielding the values of $\alpha$ and $\beta$ in Tab.~\ref{tab:cutoff}.  We observe somewhat smaller values for $\beta$ than expected, which is not too surprising given that the analytical prediction ignores multi-field effects and assumes a head on encounter with a saddle-point, both of which become good approximations for large $N$ only\footnote{Furthermore, the small bump at low $N$ (see discussion below and Fig.~\ref{fig:PNtest}) is not well fitted by a power-law, leading to a decrease of $\beta$ in a simple regression over all runs; this problem could be avoided by performing substantially more runs and including only the large $N$ tail beyond this feature.}. This is in agreement with the conclusions of Frazer and Liddle \cite{Frazer:2011br}. In addition, they observed an increase of $\beta$ if the potential becomes less featured, i.e.~by increasing the horizontal mass $m_h\equiv\varphi_i/x_i$ (we kept $m_h=M_P$ fixed throughout).

The values of $\alpha$ show a suppression with increasing $D$, which is expected, since it becomes less likely to encounter a flat stretch in field-space as $D$ is increased. This feature was also observed in \cite{Frazer:2011br} for $D=1,\dots,6$.

\begin{table}[tb]
\caption{Following the evolution in many potentials of the type \ref{potential} numerically and computing $N$ according to the algorithm in Sec.~\ref{sec:feasibility}, we plot the probability of finding $N$ e-folds of expansion in Fig.~\ref{fig:PNtest}. A power-law fit to $P(N)=\alpha N^{-\beta}$ is performed; numerical values of $\beta$ are somewhat small than the theoretical expectation $\beta=3$ in Sec.~\ref{sec:saddle}. $\alpha$ shows the expected strong suppression as $D$ is increased.}
\begin{center}
\begin{tabular}{|c|cc|cc|}
\hline
Dimension & $\#$ of runs & cut-off $\sigma_{min}$& $\alpha$&$\beta$ \\
\hline
$1$ & $20000$ & -- & $0.8$ & $2.6$ \\
$2$ & $20000$ & $0.14$ & $0.3$ & $2.3$ \\
$3$ & $20000$ & $0.14$  & $0.1$ & $2.9$ \\
\hline\hline
\end{tabular}
\label{tab:cutoff}
\end{center}
\end{table}

A novel feature not discussed in \cite{Frazer:2011tg,Frazer:2011br} is the presence of a small bump in Fig.~\ref{fig:PNtest} at small $N\sim \mbox{few}$. This feature is caused by a characteristic scale in the potential set by the modes with the lowest $J$; their effect can be understood qualitatively in one dimension: consider a single cosine as the potential, with unit amplitude and offset, $V=1+\cos(x)$ (i.e.~the $J=1$ mode only, with all other modes lumped into a constant); a randomly chosen initial value will neither be exactly at $V=0$ nor at $V=2$. Thus, the probability of finding exactly zero or infinite amount of expansion is zero. Since $P$ is continuous, there must be a maximum somewhere, which, since there is no other characteristic scale in the problem, should be near $N\sim \mathcal{O}(1)$. The presence of more modes diminishes this effect, as evident in Fig.~\ref{fig:PNtest}; however, since we strongly suppressed high frequency modes in the potential by choosing the variance according to (\ref{variance1}), the bump prevails as $D$ is increased from $1$ to $3$. Thus, we expect this feature to be characteristic for our type of ``random'' potential, but not a general prediction for random potentials.

For $D=3$, only $25$ in $10000$ runs ended up in a minimum above $V=0$ compared to $324$ and $1504$ for $D=2$ and $D=1$ respectively (we only count runs that start with $V>0$). 
Even though this is expected based on our analytic prediction that most minima have negative values of the potential as $D$ increases, $\bar{V}_p<0$ in (\ref{Vp}) (see (\ref{conjecturedprobability}) with (\ref{cana}) as well as Tab.~\ref{tab:Pmin} and Tab.~\ref{tab:Vp}), we would like to stress the importance of this result, since a negative cosmological constant is in contradiction with observations. 

Observables in runs with more than fifty-five e-folds of inflation, i.e.~in runs compatible with our universe, would be straightforward to compute; however, since almost all runs end up in an ADS vacuum (incompatible with observations), and our toy potential was not theoretically motivated anyhow, we refrain from following this avenue. If one ignores the negative cosmological constant, one may generate enough runs with sufficient inflation for $D\sim \mbox{few}$ (with modest computational effort) to provide statistics on observables such as the scalar spectral index, the tensor to scalar ratio or Non-Gausianities (NG); this was done in \cite{Frazer:2011tg,Frazer:2011br} for $D\leq 6$: as expected, since almost all acceptable runs include inflation of the slow roll type, $n_s$ and $r$ are usually compatible with observations (in the sense that $|n_s-1|\ll 1$ and small, but potentially observable $r$). Frazer and Liddle observe a dependence on the horizontal mass $m_h$ (less spread for increased $m_h$), but essentially no dependence on the dimensionality of field-space $D$. This is again expected, since the inflationary regime, taking place near a saddle-point, can usually be described by an effective-field-theory of much lower dimensionality. Consequently, NG should also be small, with two caveats: first, NG can be amplified temporarily by a sharp turn in field-space (such turns are rare for the smooth potentials under consideration); second,  as $D$ increases, it becomes less likely that an adiabatic regime had been entered at the time of horizon crossing, rendering multi-field modes essentially unpredictable with regard to NG (they continue to evolve \cite{Battefeld:2009ym,Elliston:2011dr}\footnote{At worst, one needs to follow their evolution up until the universe is reheated. Unfortunately we lack a proper understanding of (p)reheating in most multi-field models.}). Nevertheless, the expectation of commonly suppressed NG based on the horizon crossing approximation is consistent with the full numerical results in \cite{Frazer:2011br}.  

To conclude, even though dynamics in random multi-field potentials can be complicated, the overall features, such as the likelihood of inflation or the vacuum state after inflation, can be inferred from simple arguments in random matrix theory: as $D$ increases, inflation becomes increasingly unlikely, is of short duration if it occurs, and ends up most likely in an ADS vacuum if the potential is bounded and symmetric around zero. Consequences of the last observation are wide reaching, as we shall see in the next section. Further, if inflation takes place, the inflationary regime can usually be described by an effective-field-theory of low dimensionality, with all of the observational consequences this feature entails (non-Gaussianities are an exception, since a prediction for i.e.~CMBR fluctuations requires the knowledge of the evolution of the entire field content until the universe entered an adiabatic regime).

\section{Comments on the Vacuum State after Inflation \label{sec:cc}}
\subsection{Is Multi-Field Inflation unlikely?}
After inflation near a saddle-point in a higher dimensional landscape, the fields evolve rapidly since the encounter with another saddle is unlikely \cite{Aazami:2005jf}. We saw in Sec.~\ref{sec:softboundpot} at the example of a particular softly bounded potential that minima are more common once $\bar{V}\sim \bar{V}_p\ll -\sigma$, see eq.~(\ref{Vp}), that is, once the height of the potential is at a rare low value. Dynamics drive fields to such rare locations, as we saw in Sec.~\ref{sec:minima} and \ref{sec:inflation}. We argued that this feature is generic for all softly bounded random potentials. The distribution of minima, and thus for values of the cosmological constant, is far from even around zero, opposite to the assumption usually made to address the cosmological constant problem by means of anthropics coupled with eternal inflation, see i.e.~the review \cite{Bousso:2012dk}.  If the value of the (metastable) vacuum after inflation is to be positive, as in our universe, we are forced to conclude one of the following:
\begin{itemize}
\item Directly after inflation, the effective dimensionality of field-space is low so that an exponential suppression of the probability to encounter a minimum is absent; as a consequence, a landscape with an equal likelihood of positive and negative potential values can still be consistent with a small but positive cosmological constant. This requires the decoupling of almost all (massive) moduli, which may be reasonable as argued in \cite{Chen:2011ac}.
\item If the effective dimensionality of field-space remains large after inflation, the most likely vacuum to be reached dynamically is at rare, low values of the potential. For a positive cosmological constant to be likely, the potential on the landscape needs to be strongly yet precisely  shifted towards positive values (or sharply bounded from below at $V=0$).
\end{itemize}
The second option requires our toy random potential in (\ref{potential}) to be shifted upward by many standard deviations, of order $\sigma 2^{D-1}\ln(2)$. Such a potential does not appear to be natural in string theory.
We are thus left with the first option, which in turn raises another question:

\vspace{0.3cm}
\begin{center}
 \emph{If the effective dimensionality of field-space directly after inflation is low, why should it be large during/before inflation?} 
\end{center}
\vspace{0.3cm}
It appears highly fine tuned that after the (rare) encounter with a saddle-point (as needed for inflation in a higher dimensional random potential) the dimensionality of the landscape suddenly changes from $D\gg 1$ to $D_{eff}\sim \mbox{few}$. This is akin to running into a four star hotel in the Amazonian jungle (after being randomly dropped in said jungle), and after a relaxing holiday, one finds a convenient highway leading straight back to ones' porch (without encountering any junctions etc.). Without invoking anthropics (see Sec.~\ref{sec:anthropics}), this needed tuning is a strong indication that only a few light fields are dynamically relevant during inflation\footnote{The resulting presence of many, necessarily heavier moduli fields during inflation can still lead to interesting observational signatures, see i.e.~\cite{Achucarro:2010jv,Achucarro:2010da,Cespedes:2012hu,Achucarro:2012sm,Chen:2009zp,Chen:2011zf,Chen:2011tu,Jackson:2010cw,Jackson:2011qg,Jackson:2012fu}.}; hence, even though multi-field models appear to be natural in string theory with its plethora of potentially dynamical moduli fields, branes and fluxes, such scenarios are strongly disfavoured by the observation of a positive cosmological constant. A much more compelling scenario uses just one light field, so that an effective-field-theory can be applied  \cite{Burgess:2003zw,Cheung:2007st,Weinberg:2008hq}, (see \cite{Cheung:2007sv,Senatore:2009gt,Senatore:2009cf,Senatore:2010jy,Bartolo:2010bj,Bartolo:2010di,Achucarro:2012sm} for selected applications)\footnote{By single-field EFT we mean that at any point on the inflationary trajectory only one direction has a light mass; the resulting trajectory can still be strongly curved (the corresponding potential would need to contain a gorge along which the inflatons are channelled). This notion of EFT is more general than the static definition of heavy and light fields used in \cite{Avgoustidis:2012yc}, and still allows for the observational signatures in \cite{Achucarro:2010jv,Achucarro:2010da,Cespedes:2012hu,Achucarro:2012sm,Chen:2009zp,Chen:2011zf,Chen:2011tu,Jackson:2010cw,Jackson:2011qg,Jackson:2012fu} that are absent in the treatment of \cite{Avgoustidis:2012yc}.}.

Our conclusions are based on the purely classical evolution of scalar fields, invoking conservative upper bounds for the probability of encountering a minimum. The inclusion of tunnelling would strengthen our claims, since the added mobility \cite{Huang:2008jr} makes trapping in a metastable de-Sitter vacuum even less likely (the effect of resonant tunnelling \cite{Tye:2009rb} and giant leaps  \cite{ Brown:2010bc,Brown:2010mf,Brown:2010mg} improves mobility greatly as $D$ is increased).

However, our inability to reconcile the smallness of the cosmological constant with expectations from quantum-field theory is a strong indicator that we are still missing some crucial physics and we should be cautious to make such bold claims; hence, we do not advocate to throw out entirely  multi-field models based on the above reasoning; nevertheless, to explain why we got stranded at such a peculiar vacuum after inflation is an {\emph{additional}} challenge that needs to be addressed.

It should be noted that the random potentials we focussed on show the same degree of ``randomness'' in all directions, opposite to landscapes leading to a meandering inflaton, which arises if a small scale random potential is overlaid onto a smooth slow-roll slope, as in \cite{Tye:2009ff,Tye:2008ef}; for the latter our conclusions don't apply. 

Further, multi-field models with simple, similar potentials for all fields, such as in inflation driven by axions \cite{Dimopoulos:2005ac}), tachyons \cite{Piao:2002vf,Majumdar:2003kd}, M5-branes \cite{Becker:2005sg,Ashoorioon:2006wc,Ashoorioon:2008qr} or D-branes \cite{Cline:2005ty,Ashoorioon:2009wa,Ashoorioon:2009sr,Battefeld:2010rf,Firouzjahi:2010ga} etc., are not affected by our conclusions either. However, these assisted inflation scenarios \cite{Liddle:1998jc,Malik:1998gy,Kanti:1999vt,Kanti:1999ie} are not the result of sampling all possible models in string theory, but a consequence of a deliberate choice by model builders. Thus, even though these models are concrete and predictive, they tell us little about the models and predictions that are likely in string theory. More complicated setups, for which our conclusions are relevant, seem to arise naturally in string theory.  

Lastly, we would like to comment on the motivation to invoke several light fields during inflation: all current observations are fully consistent with the simplest models of single-field inflation. Concretely, recent progress on the inverse problem of reconstructing inflationary dynamics from the data by means of Baysian statistics \cite{Mortonson:2010er,Easther:2011yq,Norena:2012rs} (see also \cite{Martin:2010hh,Hazra:2012yn,  Easson:2010uw,Easson:2010zy} for general model discrimination), requires only two non-zero slow roll parameters \cite{Norena:2012rs}, and a simple single-field potential suffices. If primordial non-Gaussianities were observed, we would be forced to include more complicated dynamics, such as a sudden turn or a temporary violation of slow roll among many other possibilities; however, most of these mechanisms do not require any additional light fields.          

\subsection{A Way out via Anthropics? \label{sec:anthropics}}
It appears conceivable to evade the conclusions of the last section by means of anthropics in an eternally inflating universe  (see \cite{Polchinski:2006gy,Bousso:2012dk} for reviews). How would such a resolution look like? 

First, one employs anthropic bounds for both the upper \cite{Weinberg:1987dv} and lower \cite{Bousso:2008bu} values of the cosmological constant (CC). If $\sigma$ is much bigger than the range of anthropically allowed values of the CC, the probability distribution in (\ref{conjecturedprobability}) is essentially flat in this interval (in our concrete toy model the variance $\sigma$ in (\ref{varianceV}) increases sharply with $D$, in accord with this argument), even though it is not flat over the interval of all possible values;  hence, ending up in a deSitter vacuum in this interval is roughly as (un)likely as finding oneself in an ADS one, which is encouraging. 

But how did we get there? Anthropic bounds on negative spatial curvature as well as observations of a nearly scale invariant scalar power-spectrum require around $60$ e-folds of slow roll inflation \cite{Freivogel:2005vv}, so our universe must have encountered a saddle-point or extremely shallow region in its past. The hight of this saddle needs to be large enough to allow for subsequent reheating and nucleosynthesis, many orders above the current value of the CC \cite{Bousso:2012dk}.  

These events, finding such a saddle-point, having more than $\sim 60$ e-folds of inflation and ending up in a low vacuum with positive energy, are extremely unlikely, the more so the larger $D$ is. Thus, a mechanism is needed to sample many trajectories, such as eternal inflation. Since minima are rare for large $V$, it appears reasonable that eternal inflation is driven by fields high up in the potential (kept there by quantum fluctuations); however, our universe could also arise due to fields being trapped in a metastable vacuum prior to slow roll inflation -- after all, once started, eternal inflation populates all vacua \cite{Brown:2011ry}. The latter yields an open universe after tunnelling \cite{Coleman:1980aw,Gott:1982zf,Ratra:1994dm}, leading to the already mentioned anthropic lower bound on $N$ \cite{Freivogel:2005vv}.

Once eternal inflation is invoked, we run into a sever problem: all statistical arguments, such as the ones just made, become meaningless due to our inability to impose a mathematically well defined measure on countable infinite sets \cite{Olum:2012bn} (see also \cite{Schiffrin:2012zf}).  A possible way out of this \emph{measure problem} is to realize that eternal inflation is not past \cite{Borde:2001nh} but future eternal. Thus, at any given ``time'' the set of bubble universes in the multiverse is actually finite\footnote{The common reason infinities arise in eternal inflation (and thus a measure is sought after), is the comparison of all possible observers, future and past alike; however, such an approach violates causality: any predictions for our universe can only depend on events in the past, not possible events in the future. It appears that this fundamental principle of physics is contested in this field of research.}. Then the conceptual problems discussed in \cite{Olum:2012bn} become more practical ones: how should we define a gauge invariant time in an inhomogeneous multiverse so that we only deal with finite sets, and at what time should we evaluate probabilities (taking the limit of time going to infinity is not appropriate)? The simplest choice, taking constant proper time slices, leads to the youngness paradox \cite{Tegmark:2004qd}. Other proposals, such as using scale factor time  \cite{DeSimone:2008bq,DeSimone:2008if,Guth:2011ie} (a global measure) or the causal diamond measure \cite{Bousso:2006ge} (a local one), seem to work better (see \cite{Freivogel:2011eg} for a recent review of both global and local measures)\footnote{We would like to point out that we disagree with some of the conclusions in this review, i.e.~the ``prediction'' of the end of times \cite{Bousso:2010yn}.}. Particularly, the causal diamond measure apparently leads to the correct post-diction of the CC \cite{Bousso:2006ev,Bousso:2007kq,Bousso:2010zi}. Nevertheless, how do we know which measure is the right one in our multiverse? In the end, it is our choice of slicing that determines observables, so in what way are we predicting anything? This unfortunate freedom of choice is present due to our (conceptual) inability to probe the multiverse. Maybe all we can hope for is to follow a pragmatic approach, choose a measure ad hoc such that some observables come out right and then make predictions for others.

Following this line of thought, one needs to be aware that the entire inflationary dynamics appears to be set by anthropics. The resulting history of our universe, eternal inflation, the unlikely encounter with a reasonably hight saddle-point, inducing an unlikely long phase of slow roll inflation, followed by the extremely unlikely encounter with a metastable, low deSitter vacuum, is far from generic. 

As a consequence, studies that sample a random landscape, such as \cite{Frazer:2011tg,Frazer:2011br}, with the goal to extract distributions of observables (the scalar spectral index, the tensor to scalar ratio, non-Gaussianities etc.) are problematic if not all observational bounds are properly imposed. For example, in \cite{Frazer:2011tg,Frazer:2011br} trajectories were considered regardless of 
the value of the CC. It is not clear to us that the presence of a metastable deSitter vacuum very close to the inflationary saddle-point has no bearing on observables; for example, non-Gaussianities are sensitive to dynamics until the universe reheated and an adiabatic regime is entered \cite{Battefeld:2009ym,Elliston:2011dr}.

One conclusion is clear to us: if the landscape in string theory is as complex as it appears to be (at the energy scales relevant for inflation), our universe's history is anything but generic.

\section{Conclusion \label{sec:conclusion}}

We investigated the feasibility of softly bounded, random potentials in higher dimensional field-spaces for inflationary model building. Based on random matrix theory, we found saddle-points to be far more generic than minima or maxima for common values of the potential if $D>\mbox{few}$; as a consequence, inflation takes place most likely near a saddle, if it occurs at all. Further, shorter bursts of inflation are far more likely than longer ones, $P\propto N^{-\beta}$ with $\beta\sim 3$. We tested these general expectations in a simple toy-landscape, chosen to be reasonably smooth while remaining softly bounded by means of a Gaussian factor ($P\propto \exp(-V^2/2\sigma_V^2) /\sigma_V$), and we found good agreement.

We followed with theoretical and numerical investigations of the final state after inflation in such a random potential. Since the probability of running into a minimum decreases dramatically for positive values of the potential as the dimensionality of field-space is increased, see eqn.~(\ref{conjecturedprobability}), it becomes exceedingly unlikely to end up in a universe with a positive cosmological constant after inflation. We confirmed this expectation numerically for the same toy-landscape; for instance, in the presence of three fields less than 25 in 10000 runs (indiscriminate of whether they include inflation or not) ended up in a minimum above zero. Taken together with the observational requirement of around sixty e-folds of inflation, it is extremely unlikely that our universe originated from dynamics involving more than a couple of light fields during inflation. To reach this conclusion, we used conservative estimates throughout and we deliberately avoided anthropic arguments, which could be used as a way out. 

As a consequence, the effective-field-theory of single-field inflation in the presence of some heavier fields, see i.e.~\cite{Cheung:2007st,Weinberg:2008hq,Achucarro:2012sm,Jackson:2010cw,Jackson:2011qg,Jackson:2012fu} for recent progress, could not only be entirely sufficient to tell us everything we will ever know about inflation, but appears to be favoured over more complex multi-field models in light of a positive cosmological constant.

\acknowledgments
T.B.~and D.B.~are grateful for hospitality at the Institude de AstroParticule et Cosmologie (APC, Paris) and the University of Portsmouth. We would like to thank C.~Behrens, J.~F.~Dufaux, R.~Easther, J.~Frazer, T.~Giblin, L.~McAllister, J.~Niemeyer and S.~Patil for discussions.

\end{document}